\documentclass[aps,prl,twocolumn,nobibnotes]{revtex4}
\usepackage{amsmath,amssymb}
\usepackage{color, bm,graphicx}
 
\newcommand{\be}{\begin{equation}}
\newcommand{\ee}{\end{equation}}

\newcommand{\irr}{\text{irr}}

\newcommand{\G}{\Gamma^{(n)}}
\renewcommand{\k}{{\underline k}}

\newcommand{\p}{{\underline p}}

\begin{document}

\title{Self-Similarity of Loop Amplitudes}
\author{Kang-Sin Choi}
\email{kangsin@ewha.ac.kr}
\affiliation{Scranton Honors Program, Ewha Womans University, Seoul 03760, Korea} 
\affiliation{Institute of Mathematical Sciences, Ewha Womans University, Seoul 03760, Korea}

\begin{abstract}
We present an amplitude-generating formula in renormalizable quantum field theory. It reflects the self-similarity of loop amplitudes, in which an amplitude can also be a subamplitude of another. Amplitudes are generated by a small number of ``irreducible'' ones, which may replace tree-level couplings to form more complex amplitudes.
\end{abstract}

\maketitle

Quantum Field Theory provides a successful framework for describing fundamental interactions. As a Lorentz covariant tool for $S$-matrix calculations, it employs loop amplitudes of quantum nature \cite{Nambu:1968rr, Coleman:1973jx}. Renormalizability is one guideline for constructing theory, which restricts the possible interactions, ensuring their relevance at low energies and preserving unitarity \cite{Weinberg:1979pi}.

Calculation of high loop-order amplitudes is important, e.g., in the precision test \cite{Aoyama:2012wk, Aoyama:2017uqe}.
While Feynman rules structure these calculations, they introduce significant redundancies. These redundancies arise from the self-similarity of loop amplitudes, which allow amplitudes to be embedded as subamplitudes within larger ones.

\begin{figure}[t]
\begin{center}
\includegraphics[scale=0.53]{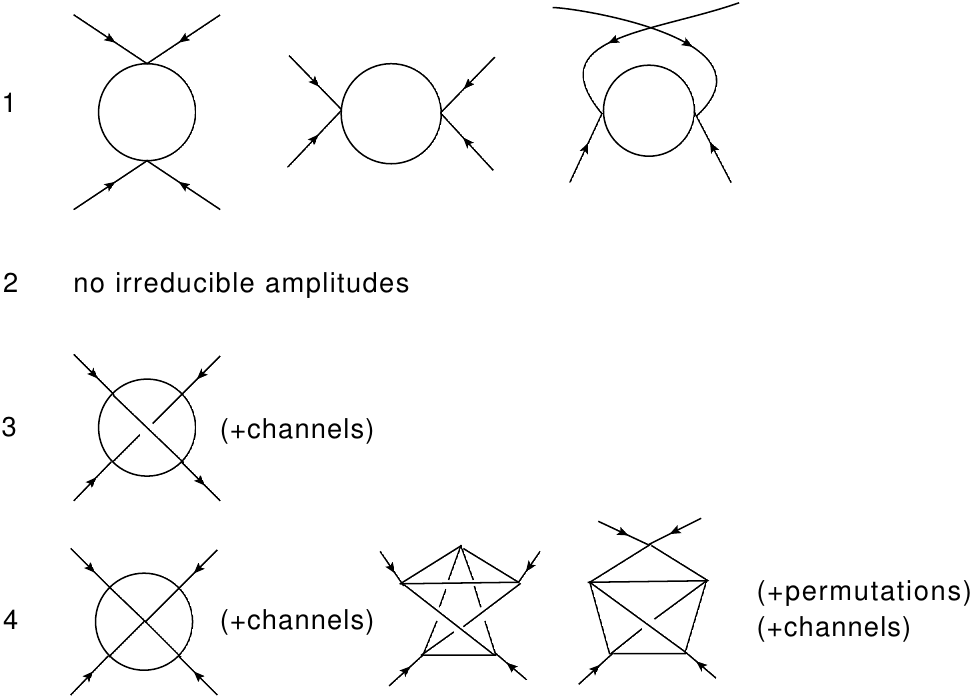}
\end{center}
\vskip -0.5cm
\caption{Irreducible dirgrams in $\phi^4$-theory up to four-loop order. They are primary elements of multiloop amplitudes. Any subloop is not $r$-contractable. At each loop, we have different types of diagrams with channels.  \label{figIr}}
\vskip -0.48cm
\end{figure}

We present a simple but powerful formula for removing such redundancy \cite{Choi:2025tus}. Namely, all the amplitudes are obtained from lower-loop amplitudes by replacing the couplings with some ``irreducible'' amplitude. This provides an efficient method for amplitude generation (see also \cite{Choi:2024hkd, Choi:2024cbs, Kleinert:1982ki, Kastening:1999fy, Kleinert:1999uv, Ilderton:2005vg, Choi:2007nb, Kim:2008hda, Abe:2009dr, Borinsky:2022lds}). For instance, in $\phi^4$-theory, solely the three one-loop irreducible amplitudes in Fig. \ref{figIr} generate 24, 312, 5616, 129168 four-point amplitudes at 2,3,4,5 loops, respectively. Up to order 4, we have only a handful of such elementary amplitudes as in Fig. \ref{figIr} proliferating higher-order ones. Although many of them are the same, we can systematically generate complete amplitudes without missing them.

What we observe via the $S$-matrix is expanded by vertex functions, or one-particle-irreducible (1PI) $n$-point amplitudes \cite{Jackiw:1974cv,Choi:2025tus}
\be  \label{vertexfn}
 \Gamma^{(n)} (\p) = \langle \phi_1(p_1) \phi_2(p_2) \dots \phi_n(p_n) \rangle_{\text{1PI}},
\ee
with momentum conservation $\sum_{i=1}^n p_i = 0$. They admit expansion in the loop order \cite{Nambu:1968rr,Coleman:1973jx,Choi:2025tus}
\be \label{effcoup}
 \G(\p) = g^{(n)} + \sum_{l=1}^{\infty} \G_l(\p), \quad \G_l(\p) \propto \hbar^l. 
\ee

We show that the vertex functions at $L$-loop are generated as 
\be \label{recform}
 \G_{L} (\p) = \sum_{m=0}^{L-1}  \sum_{v_a,l_{ab}}  S_{v,l}^{-1} \G_{m} (\p) \Big|^1_{ g^{(n_a)} \to \Gamma_{L-m}^{(n_a),\irr}} . 
\ee
It is the sum of {\em formally} $m$-loop ``base'' amplitudes $\G_{m} (\p)$, with {\em one} of the coupling $g^{(n_a)}$ replaced by an order-$(L-m)$ amplitude $\Gamma_{L-m}^{(n_a),\irr}(\p,\k)$. This replacement is justified by the expansion structure (\ref{effcoup}).

We want to produce the coupling systematically without duplicates; we use the ``irreducible'' one. For this, we first define {\em $r$-contraction:} we replace a (sub)amplitude $A^{(n_a)}_l$ to the renormalized tree coupling $g^{(n_a)}$. In the diagram, we literally contract the corresponding renormalizable subamplitude. A necessary condition is that there must exist the corresponding coupling in the Lagrangian. We consider a renormalizable theory. 

For a given $r$-contractable diagram, a successive $r$-contraction of subamplitudes, in the end, gives us an {\em irreducible} amplitude. If no proper subamplitude is $r$-contractable, we define it irreducible. We also define a one-loop amplitude irreducible. Irreducible amplitudes may have different channels having the same external legs. 

If one amplitude can replace a coupling, a sum of them can, too.
We define an {\em irreducible effective coupling} as the sum of all the irreducible amplitudes of the same order
\be
  \Gamma^{(n),\irr}_l(\p) \equiv \sum_{\text{types}} \sum_{\text{channels}} A^{(n)}_l(\p), 
\ee
where
\be
  A^{(n)}_l(\p) = \int d^4k_1 \dots d^4k_l \prod_{v_a} g_a P_a \prod_{l_{ab}} \Delta_{ab}(\p,\k),
\ee
with standard normalization. Normally, we calculate it order by order using Feynman rules \cite{Jackiw:1974cv}. At each vertex $v_a$, we have a coupling $g_a$ and a polynomial $P_a$ in the internal and inflowing momentum. At each line $l_{ab}$, we have a propagator $\Delta_{ab}$.  

For instance, in $\phi^4$-theory, at one-loop order, there are quartic amplitudes with the diagrams in Fig. \ref{figIr}, in three channels. In this case $\Gamma^{(4)}_1(\p) = \Gamma^{(4),\text{irr}}_1(\p)$, as in (\ref{lambdacorr}) below, since all of them are irreducible. At loop order 3, there is only one type of irreducible amplitudes $\Gamma^{(4),\irr}_3(\p)$ in three channels as in Fig. \ref{figIr}. In contrast, the reducible amplitudes in $\Gamma^{(4)}_3(\p)$ are generated from two-loops. At loop order 4, we have two different types of amplitudes in Fig. \ref{figIr} contributing to $\Gamma^{(4),\irr}_4(\p)$.

Now, the {\em resolution} $g^{(n_a)} \to \Gamma_{L-m}^{(n_a)} (\p,\k)$, done in (\ref{recform}), is the reverse of $r$-contraction; we replace a single coupling $g^{(n_a)}$ on one vertex $a$ with the irreducible effective coupling $\Gamma_{L-m}^{(n_a)} (\p,\k),$ now depending on the part of the internal momenta $k_1,\dots, k_{L-m}$. The latter inherits the conserved momenta at the former \cite{Zimmermann:1969jj}. We do the same to a propagator with $\Delta_{ab}(\p,\k)=i/\Gamma_{L-m}^{(2)}(\p,\k)$ along one line $l_{ab}$. 

The construction is inductive, which grows the amplitude by inserting an irreducible amplitude one by one. 
Start with $n$-point couplings $g^{(n)} \equiv \G_0$. Assume we have all the base $n$-point couplings $\Gamma_l^{(n)},$ of order $l=0,\dots L-1$, for all $n$. Now the amplitude (\ref{recform}) is constructed from the resolution by an {\em irreducible} amplitude. 

This generates the complete set of amplitudes of the loop order $L$. The irreducible amplitude contains no further $r$-contractable subamplitude, so it is the minimal building block in the resolution. Therefore, the above procedure exhausts the diagram up to the order $(L-1)$. In other words, there is no intermediate amplitude under resolutions. The case $m=0$ is its own coupling, and we have to find irreducible amplitudes of order $L$. 

In front of the base amplitudes, we put the symmetry factor $S_v^{-1}$ or $S_l^{-1}$, if we have different vertices or lines giving the same diagram by resolution. Once we fix it, the rest of the asymmetric diagrams acquire correct multiplicities. This plays the same role as the factorial in the Dyson series, reducing the number of identical diagrams, and we have correct multiplicity after time ordering in $\G$ \cite{Dyson:1949ha}. Different channels of an irreducible amplitude may be distinguished, but if they become the same after being used in the resolution, the above factor takes care of the repetition. In Fig. \ref{figM2}, in the $s$-channel base, the insertion of $t$ and $u$ channel amplitudes give the same diagram, and it is reduced by the symmetry factor used in the insertion of the $s$-channel.

The self-similar structure is tied to the renormalizability of the theory. While non-renormalizable operators cannot be systematically classified, renormalizable operators have a limited number of external legs due to Weinberg's theorem \cite{Weinberg:1959nj}; for instance, in $\phi^4$-theory, this number cannot exceed four.
High-order loop diagrams maintain limited patterns because they, too, are constructed from renormalizable couplings. As loop order increases, lines grow more numerous than vertices, forcing them to form $r$-contractable loops. Consequently, complex amplitudes are naturally constructed from irreducible amplitudes as essential building blocks.

\begin{widetext}


\begin{figure}[t]
\begin{center}
\includegraphics[scale=0.75]{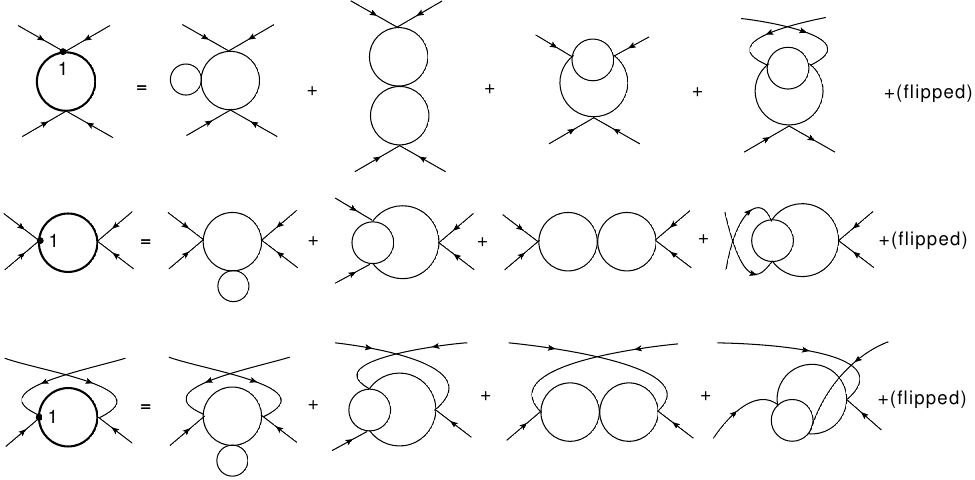}
\end{center}

\caption{Two-loop corrections for the quartic coupling in $\phi^4$ theory, generated by formally one-loop amplitude. We obtain them by replacing one of the vertices or lines in a one-loop diagram ($s,t,u$ respectively, rows) with the one-loop renormalized $\G_{1}(\p)$ $s,t,u$-vertices (columns). We have one more set of flipped ones. Some are redundant and give the correct symmetry factors. \label{figM2}}
\end{figure}

As examples, we generate quartic couplings in $\phi^4$-theory with the coupling $g$ to four-loop order. We have usual one-loop couplings in three channels in the Mandelstam variables \cite{Choi:2025tus}
\be  \label{lambdacorr}
i \Gamma^{(4)}_{1}(\p) = \sum_{p^2=s,t,u} \frac{1}{2} \int \frac{d^4k}{(2\pi)^4} (-ig)^2 D_0((p+k)^2) D_0(k^2) \equiv i \Gamma^{(4),\irr}_{1}(\p) ,
\ee
which are irreducible, too. We can generate propagator $i/\Gamma^{(2)}$ in the same way. 

The two-loop corrections to the quartic coupling are the one-loop amplitudes $\Gamma^{(4)}_{1}(\p)$ with one vertex replaced by the above one-loop irreducible vertex $\Gamma^{(4),\irr}_{1}(\p)$,
\be \label{FormalQuartic} 
\begin{split}
i \Gamma^{(4)}_{2}(\p) & =  \frac{1}{2} \sum_{\text{2 vertices}} i \Gamma^{(4)}_{1}(\p) \Big|^1_{g^{(4_a)} \to \Gamma^{(4_a),\irr}_{1} } + \frac{1}{2} \sum_{\text{2 lines}}^4 i \Gamma^{(4)}_{1}(\p) \Big|^1_{g^{(2_a)} \to \Gamma^{(2_a),\irr}_{1} }  \\
 &= \frac{1}{4} \int \frac{d^4k}{(2\pi)^4} [- i g] 
 D_{0}((p+k)^2) D_{0}(k^2)[-i \Gamma^{(4)}_{1}(\p,k)]  
 + \frac{1}{4} \int \frac{d^4k}{(2\pi)^4} [- i \Gamma^{(4)}_{1}(\p,k)]
 D_{0}((p+k)^2) D_{0}(k^2)[- i g] \\
  & \quad
  + \frac{1}{4} \int \frac{d^4k}{(2\pi)^4} [- i g] 
 D_{1}((p+k)^2) D_{0}(k^2)[- i g]
 + \frac{1}{4} \int \frac{d^4k}{(2\pi)^4} [- i g]
 D_{0}((p+k)^2) D_{1}(k^2)[- i g] 
 \end{split}
\ee
We have two vertices and two propagators, which the ones of loop order one may replace. Their equivalence in generating the same amplitudes is reflected in the factor $1/2,$ respectively. We focus on the vertex resolution here. Each vertex lies inside the loop with the internal momentum $k$, to which four momenta $(\p,k) \equiv \{-k,-p_i,-p_j,k+p_i+p_j\}$ flow in. Here, $p_i$ and $p_j$ $(i,j)=(1,2),(1,3),(1,4)$ originate from the channel of the original one-loop diagram as in (\ref{lambdacorr}). In the {\em subdiagrams}, we also have three possible channels as functions of $s'=(p_i+k)^2, t'=(p_j+k)^2, u'=(p_i+p_j)^2$, respectively:
\begin{align} 
\Gamma^{(4)}_{2}(\p)  &= g^3[ W(s) + W(t) + W(u)] , \\
\begin{split} \label{Wschannel} 
 W(s) &= \frac12 \int \frac{d^4k}{(2\pi)^4}D((p_3+p_4+k)^2)D(k^2)  \Big(  V(s)  + V((p_4+k)^2) +  V((p_3+k)^2) \Big) ,
\end{split} 
\end{align}
with similar $t$, $u$ channel amplitudes.
We have two-loop corrections as a result, whose corresponding diagrams are drawn in Fig. \ref{figM2}, where the dot and the thick lines indicate the resolved loop.  
We may check that all the amplitudes have the correct symmetric factors, noting that the last two amplitudes in (\ref{Wschannel}) are the same.
\be \begin{split}
 \int &d^4k d^4 \ell D_0((p_3+p_4+k)^2)D_0(k^2)  D_0((p_4+k+\ell)^2)D_0(\ell^2) \\
  &=  \int d^4k' d^4 \ell'  D_0(k^{\prime 2})D_0((p_3+p_4+k')^2) D_0((p_4+k'+\ell')^2) D_0(\ell^{\prime 2}),
\end{split}
\ee
by $p_3+p_4+k = - k', \ell' = -\ell.$
The last two diagrams in the first line in Fig. \ref{figM2} also explain why they are equivalent, although ones with different channels generate them. At two-loop order, there are no reducible diagrams.

Three-loop amplitudes can be generated by the resolution of the two-loop base ones by the one-loop irreducibles. There are no irreducible ones at loop order two. We also have order-three irreducible diagrams as in Fig. \ref{figIr},
\be\begin{split}
 A_3^{(4)}(\p) = g^4  \int \frac{d^4k_1}{(2\pi)^4}\frac{d^4k_2}{(2\pi)^4}&\frac{d^4k_3}{(2\pi)^4}D_0(k_1^2)D_0((k_1+p_1-k_2)^2)D_0(k_2^2) \\
 & \quad \times D_0((k_2+p_2-k_3)^2)D_0(k_3^2) D_0((k_3-k_2+p_1+p_4)^2),
 \end{split}
\ee
plus channel variants \cite{Choi:2025tus}. In terms of diagrams, we have 
\begin{center}
\includegraphics[scale=0.6]{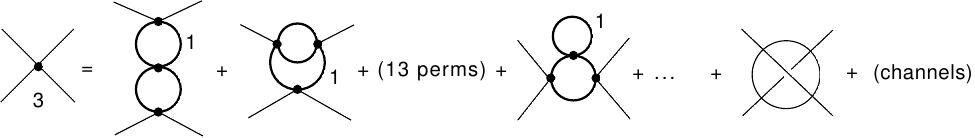}
\end{center}
\vskip -0.2cm
The amplitude is
\be
i \Gamma^{(4)}_{3}(\p)  = \frac{1}{3} \sum_{\text{3 vertices}} i  \Gamma^{(4)}_{2}(\p) \Big|^1_{g^{(4_a)} \to \Gamma^{{(4_a)},\irr}_{1} } +\frac{1}{4} \sum_{\text{4 lines}} i  \Gamma^{(4)}_{2}(\p) \Big|^1_{g^{(2_a)} \to \Gamma^{{(2_a)},\irr}_{1} }   + i  \Gamma^{(4),\irr}_3 (\p).
\ee 
The symmetry factor comes from 3 vertices or 4 lines, respectively.

Finally, four-loop amplitudes are formed by formal one- and three-loop reducible amplitudes as bases because there are irreducible amplitudes at orders one and three, as shown in Fig. \ref{figIr}. In addition, we have three types of order-four irreducible ones in various channels
\be
 \Gamma^{(4),\irr}_4(\p) = \sum_{\text{3 types, perms}} \sum_{\text{channels}} A^{(4)}_4(\p).
\ee 
Therefore, we have
\begin{center}
\includegraphics[scale=0.7]{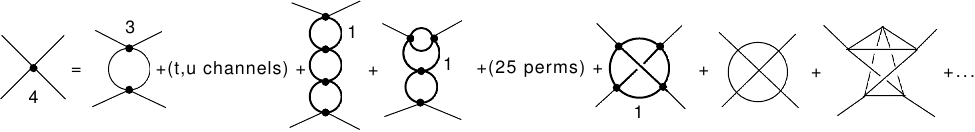}
\end{center}
\vskip -0.2cm

\be \begin{split}
i \Gamma^{(4)}_4(\p)  &= \frac{1}{2}  \sum_{\text{2 vertices}} i\Gamma^{(4)}_{1}(\p) \Big|^1_{g^{(4_a)} \to \Gamma^{{(4_a)},\irr}_{3} }  + \frac{1}{4} \sum_{\text{4 vertices}}i \Gamma^{(4)}_{3}(\p) \Big|^1_{g^{(4_a)} \to \Gamma^{{(4_a)},\irr}_{1} } \\
& \quad+ \frac{1}{6} \sum_{\text{6 lines}}i \Gamma^{(4)}_{3}(\p) \Big|^1_{g^{(2_a)} \to \Gamma^{{(2_a)},\irr}_{1} } + i\Gamma^{(4),\irr}_4 (\p).
\end{split}
\ee

\begin{figure}[t]
\begin{center}
\includegraphics[scale=0.75]{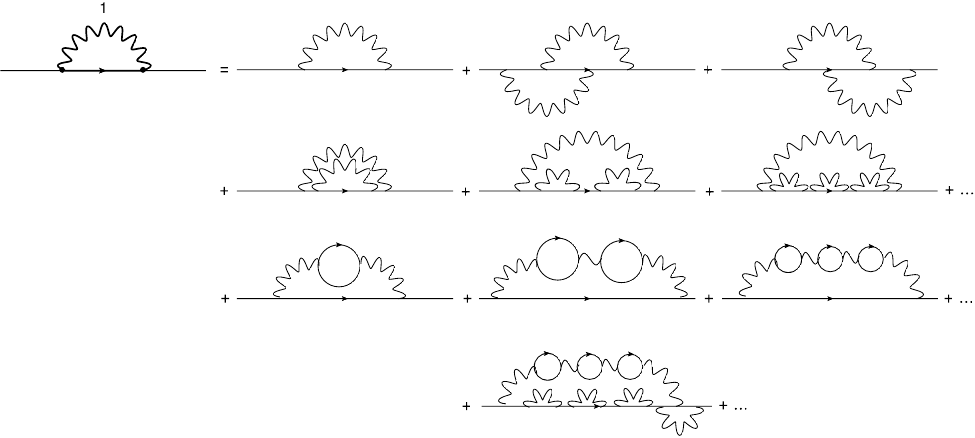}
\end{center}
\vskip -0.5cm
\caption{Generation of electron self-energy loop amplitudes from a weighted one-loop amplitude in QED. In the propagator, we may have one-particle-reducible subamplitudes. \label{fig:S1}}
\end{figure}

\end{widetext}

The above generation method is readily generalized to other renormalizable theories since the notion of $r$-contraction, irreducibility and hence the resolution remains valid. For example, we can also generate general amplitudes in quantum electrodynamics (QED) similarly, as in Fig. \ref{fig:S1}, by calculating the vertex functions now with electrons and photons.

The above construction enables us to understand perturbative quantum field theory in terms of the irreducible amplitudes and their chains. This method is also powerful in renormalizing amplitudes: we may always use the irreducible amplitudes that are renormalized \cite{Choi:2025tus}. 

This work is partly supported by the grant RS-2023-00277184 of the National Research Foundation of Korea.

\end{document}